\documentclass[10pt,letterpaper]{article}
\usepackage{amsmath}

\usepackage{opex3}

\input{tcilatex}

\begin{document}

\title{Self ordering and superradiant backscattering of a gaseous beam in a ring
cavity with counter propagating pump}
\author{C. Maes}
\address{Department of Physics, University of Arizona, PO Box 210081\\ Tucson, AZ, 85721}
\email{maes@physics.arizona.edu}

\author{J. K. Asb\'{o}th$^{1,2}$ and H. Ritsch$^2$}
\address{$^1$Research Institute of Solid State Physics and Optics,Hungarian Academy of Sciences,H-1525 Budapest P.O. Box 49, Hungary\\$^2$Institute of Theoretical Physics, University of Innsbruck, Technikerstrasse 25, A-6020 Innsbruck, Austria}
\email{Janos.Asboth@uibk.ac.at, Helmut.Ritsch@uibk.ac.at}

\begin{abstract}
We study the threshold conditions of spatial self organization combined with
collective coherent optical backscattering of a thermal gaseous beam moving
in a high Q ring cavity with counter propagating pump. We restrict ourselves
to the limit of large detuning between the particles optical resonances and
the light field, where spontaneous emission is negligible and the particles
can be treated as polarizable point masses. Using a linear stability
analysis in the accelerated rest frame of the particles we derive an
analytic bounds for the selforganization as a function of particle number,
average velocity, temperature and resonator parameters. We check our results
by a numerical iteration procedure as well as by direct simulations of the
N-particle dynamics. Due to momentum conservation the backscattered
intensity determines the average force on the cloud, which gives the
conditions for stopping and cooling a fast molecular beam.
\end{abstract}










\section{Introduction}

Light forces have become an essential tool to cool, trap and manipulate
atomic gases\cite{Nobel cooling}. Ultimately it allowed to reach BEC and
thus the lowest temperatures experimentally available\cite{Nobel BEC}. Ever
since this success there has been an ongoing quest to extend laser cooling
to more atomic species and also molecules. Here the need for a closed
optical cycling transition and the preparation of an initial sample of
sufficient low temperature and high phase space density to start the cooling
process has posed severe limitations. In particular for molecules,
spontaneous emission induced by the laser fields populates a great manifold
of electronic and rovibrational levels, which hampers the standard laser
cooling techniques. \

Recently is has been theoretically predicted\cite{Domokos03,Beige,Chu} and
experimentally confirmed \cite{Vuletic1,Rempe,Kuhn} that using high Q
resonators for the cooling light fields, the role of spontaneous emission
can be strongly suppressed in laser cooling. As one further decisive
experiment superradiant enhancement of cavity cooling was recently
demonstrated for Cesium atoms in a confocal cavity\cite{Vuletic2}. Although
the experiments were performed on atoms, this opens very promising prospects
on molecules as well. Trapping and cooling of atoms in a far detuned ring
cavity field was recently also achieved\cite{Hemmerich,Zimmerman,elsasser04}%
. \ \ \ \ \

As most sources of gaseous cold molecules are fast beams, the second problem
to be dealt with is stopping the molecules of such a beam in a very short
distance within the vacuum chamber and trap them. A first success in this
direction was achieved by using periodic electrostatic fields to stop a
molecular beam\cite{Meijer} or by creating the molecules already cold by
laser ablation or directly from an atomic magneto optical trap\cite{Pillet}.
\ Here we study the prospects of using a ring cavity setup as an alternative
way to slow down a gaseous beam. The particles kinetic energy is converted
into light energy using inelastic coherent backscattering. Such a setup,
called correlated atomic recoil laser (CARL), was historically proposed in
analogy to the Free electron laser (FEL) as laser gain medium using atoms%
\cite{Carlold,Bonifacio}. Many theoretical investigations of these
phenomenon, sometimes also called recoil induced resonance (RIR), produced
partly controversial results and interpretations \cite{Berman92,perrin}. For
quite some time experimental efforts to observe clear signatures of CARL
were only partly successful\cite{Bigelow}.

Only recently, a first unambiguous observation of the (inverse-) CARL
mechanism to accelerate a cold gas by coherent light scattering in a cavity
was achieved in T\"{u}bingen using Cesium atoms\cite{Zimmermann,slama04}.
These results are in very good agreement with predictions from microscopic
dynamic models\cite{Gangl} as well as collective macroscopic descriptions of
atoms in ring cavities\cite{Robb}. Even stronger stopping forces were
observed and extrapolated by {Vuleti\'{c}} in a standing wave cavity setup%
\cite{Vuletic2,Vuletic3}, where a special bistable atomic selforganization
process superradiantly enhances the light forces\cite{Domokos03,Asboth05}.
In all of the experiments the particles were atoms with a closed optical
cycle, which were cold and relatively slow from the start. Nevertheless
these results show that the method has great promise.

Here we investigate the experimental conditions necessary to stop and trap a
fast molecular beam by help of a single side pumped ring cavity geometry.
The idea relates closely to the original idea of CARL, but we are interested
in stopping and cooling particles and not in laser gain. In contrast to most
previous work we consider the limit of large detuning, where inversion and
spontaneous emission from the particles plays no role. This enables one to
adiabatically eliminate the excited states from the dynamics, which allows
us to apply the results to any point like objects with suitable
polarizability and in particular cold molecules.

In this limit the equations simplify dramatically but, of course, the atom
field interaction strength per particle gets very small and only cooperative
light scattering by a large number of scatters can induce a sufficiently
strong force. \ In the quasi 1D geometry along the cavity axis this requires
a periodic particle arrangement, so that they act like a Bragg mirror
coupling the light fields within the resonator. For favorable parameters the
interference of the injected and backscattered light creates an optical
potential suitable to stabilize such a regular distribution or even induce
it via selforganization from an initial flat distribution.

Note that momentum conservation assures that the total force on the cloud is
directly related to the backscattered field intensity amounting to two
photon momenta per photon decaying from the backward mode. Hence the average
force on the gas will always point to the same direction and this
acceleration prevents the system from reaching a steady state in the long
run. Nevertheless at an intermediate timescale one can expect the atoms to
form a transient spatial equilibrium distribution with respect to their
accelerated center of mass.

In the following we try to answer two closely related questions in this
respect: (1) Under what conditions and circumstances is a homogeneous
particle distribution within the resonator unstable against small density
fluctuations so that the atoms will start to selforganize? (2) How does the
self consistent atom field distribution look like in the regime where self
ordering happens and what is the corresponding field intensity and average
force on the cloud? As a third important question we address the scaling of
the system with particle number, volume and required pump intensity to see
whether the effect persists on a macroscopic scale.

Methodically we will adapt an idea already successfully applied to
selforganization of atoms in standing wave cavity fields. For this one
iteratively determines the self consistent steady state of the cavity field
with particles ordered in a way just to create this very same field. The
iteration is done assuming fast atomic thermalization and field
equilibration but only slow change of the average atomic temperature. In
order to adapt the method to particles in a ring cavity we have to change to
the accelerated rest frame of the atomic clouds center of mass and
additionally assume that the equilibration is faster than the center of mass
acceleration. This allows us to find an accelerating position distribution
with a largely time independent shape. Investigating the stability and self
consistence of this distribution will then give us limiting conditions for
which the proposed method should work. In order to check the predictions we
then will directly simulate the microscopic equations of motions for finite
ensembles of particles and the fields.

\bigskip

\section{Model}

As model system let us consider a large number $N$ of linearly polarizable
point particles interacting with two degenerate counter propagating linearly
polarized plane wave ring cavity modes.

The electric field inside the cavity is then
\begin{equation}
\overrightarrow{E}(x,t)=\overrightarrow{E}^{(+)}(x,t)e^{-i\omega_{P}t}+%
\overrightarrow{E}^{(-)}(x,t)e^{i\omega_{P}t},
\end{equation}%
\begin{equation}
\overrightarrow{E}^{(+)}(x,t)={\Large \varepsilon}\left[
a_{+}(t)e^{ikx}+a_{-}(t)e^{-ikx}\right] \overrightarrow{e}_{z}   \label{E}
\end{equation}%
\begin{equation}
\overrightarrow{E}^{(-)}=\left[ \overrightarrow{E}^{(+)}\right] ^{\dagger}
\end{equation}
where the electric field per photon is ${\Large \varepsilon}=\sqrt{\hbar
\omega_{P}/2\epsilon_{0}V}$ and $a_{\pm}(t)$ gives the field amplitude.

The pump and cavity frequency are assumed to be sufficiently detuned from
any optical resonance so that the particles are only weakly excited and the
polarization can be adiabatically eliminated. Thus they experience a Stark
shift linear in the local intensity and act like a dynamic refractive index
in the field. Let the parameter $U_{0}$ denote the light shift per photon in
the field. From energy conservation one sees that $U_{0}$ then also has to
give the frequency shift of the cavity mode resonances per atom. In terms of
the two level linewidth $\Gamma$, detuning $\Delta_{a}$ and atom-field
coupling $g$ the effective mode frequency shift reads:
\begin{equation}
U_{0}=\frac{\Delta_{a}}{\Gamma^{2}+\Delta_{a}^{2}}g^{2}.
\end{equation}

In a typical setup one sends the molecular beam at average velocity $v_{0}$
into the ringcavity nearly antiparallel to the driven circulating field. The
molecules might be moving quite fast (up to \symbol{126}100 m/s) without
some pre-slowing so that the ratio of the Doppler shift to the linewidth $%
kv/\kappa,$ can be larger than unity but not too large. At much larger
velocities, the pump and or the backscattered fields are far of resonance,
which requires stronger and stronger pump amplitudes to generate a
noticeable force.

In the following we will shift our reference frame to coincide with the
particles center of mass motion. Hence, while a particle at rest in the
cavity would see a single cavity detuning $\Delta_{c},$ the molecular beam
experiences two different Doppler shifted cavity fields with detunings $%
\Delta_{\pm}=\Delta_{c}\pm kv.$ Thus the equation of motions for the field
mode amplitudes are\cite{Gangl}:
\begin{equation}
\frac{d}{dt}a_{\pm}(t)=\left[ i\Delta_{\pm}-iNU_{0}-\kappa\right] a_{\pm
}(t)-iNU_{0}\left\langle e^{\mp2ikx}\right\rangle a_{\mp}(t)+\eta_{\pm}.
\label{modederiv}
\end{equation}

Here $N$ denotes the particle number and $\eta_{\pm}$ gives the cavity pump
amplitudes. \ We see that the particles density distribution couples the two
counter propagating modes via the spatial average $R_{\pm}=\left\langle
e^{\mp2ikx}\right\rangle $, which is called particle bunching parameter and
one has $R_{+}=R_{-}^{\ast}.$ A uniform distribution gives $R_{\pm}\approx0$%
, while an optimally ordered phase gives $\left\vert R_{\pm}\right\vert
\rightarrow1.$

As second part in our model we include the force of the light field on the
particles. Again assuming the large detuning limit this force is dominated
by the dipole force\cite{Gangl}. This force can be derived from the
effective optical potential, which is determined by the local field
intensity simply by:
\begin{equation}
{\Large \varepsilon}^{2}V(x)=U_{0}\left\vert E(x)\right\vert ^{2}.
\label{optpot}
\end{equation}

For $U_{0}<0$, i.e. $\Delta_{a}<0$, this attracts atoms to maxima of the
cavity field. Using the definition of the field in terms of the amplitudes $%
a_{\pm},$ $E^{+}(x)={\Large \varepsilon}(a_{+}e^{ikx}+a_{-}e^{-ikx})$ we
get:
\begin{align}
\left\vert E{(}x{)}\right\vert ^{2} & ={\Large \varepsilon}^{2}\left[
\left\vert a{_{+}}\right\vert ^{2}+\left\vert a{_{-}}\right\vert
^{2}+a_{+}a_{-}^{\ast}e^{2ikx}+(a_{+}a_{-}^{\ast})^{\ast}e^{-2ikx}\right]
\label{intensity} \\
\quad & ={\Large \varepsilon}^{2}\left[ \left\vert a{_{+}}\right\vert
^{2}+\left\vert a{_{-}}\right\vert ^{2}+2\left\vert a_{+}a_{-}^{\ast
}\right\vert \cos(2kx+\alpha)\right] ,
\end{align}
with $\alpha$ being the argument of $a_{+}a_{-}^{\ast}$. Note that for
calculating the force the constant terms can be omitted from the potential
and the average force then simply reads:

\[
f(x)=-U_{0} \nabla\left\vert E{(}x{)}\right\vert ^{2} = -2\hbar
kU_{0}\left\vert a_{+}a_{-}^{\ast}\right\vert \sin(2kx+\alpha).
\]

This force will then act on the atomic distribution and in general one gets
a rather complex coupled atom field dynamics\cite{domokos02b}. Note that
neglecting spontaneous emission (no radiation pressure force) we get
momentum conservation of the cavity field modes and gas. This has the
consequence that for single side pumping, for every photon in the counter
propagating mode one has transferred a momentum of $2\hbar k$ to the cloud.
Hence in steady state the average force is directly proportional to the
output intensity of the counter propagating mode and we will always have
acceleration in the pump direction.

\section{Coupled particle field dynamics}

\subsection{Quasistationary field}

Lets us now get some first qualitative insight in the ongoing dynamics for a
thermal cloud in a single side pumped ring cavity ($\eta_{+}=\eta,\eta_{-}=0$%
). While for an empty cavity the field intensity and thus the optical
potential is constant along the axis, the particles will coherently
backscatter some light and create a periodic intensity modulation. In a
typical experimental setup this field redistribution will happen much faster
than the typical time on which the particle distribution evolves. Hence the
field will almost instantaneously reflect the momentary particle
distribution. As the only relevant quantity of the particle distribution
entering the field equations is the parameter $R_{-}=\left\langle
e^{2ikx}\right\rangle $, the field amplitudes will relax towards:

\begin{align}
a_{+} & =i\eta\frac{\Delta_{-}-NU_{0}+i\kappa}{{(}\Delta_{+}{-}NU_{0}{%
+i\kappa)(}\Delta_{-}{-}NU_{0}{+}i\kappa{)-}N^{2}U_{0}^{2}\left\vert
R_{-}\right\vert {^{2}}}, \\
\ a_{-} & =i\eta\frac{NU_{0}R_{-}}{{(}\Delta_{+}{-}NU_{0}{+i\kappa)(}%
\Delta_{-}{-}NU_{0}{+}i\kappa{)-}N^{2}U_{0}^{2}\left\vert R_{-}\right\vert {%
^{2}}}.
\end{align}

To get some more insight in the physical meaning of $R_{-}$ let us consider
the prototype case of a periodically modulated gas density with symmetric
maxima at $x=x_{0}+n\lambda/2$. $R_{-}$ than can be written as $R_{-}
=\left\vert R_{-}\right\vert e^{2ikx_{0}}$, where the modulus $\left\vert
R_{-}\right\vert $ measures the modulation depth and the location of the
maximum density determines the phase of $R_{-}$. Inserting these expressions
into the field equations, we then get:

\begin{equation}
a_{+}a_{-}^{\ast}=\left\vert {\eta}\right\vert ^{2}\frac{{(}\Delta_{-}{-}%
NU_{0}{+}i\kappa{)}NU_{0}\left\vert R_{-}\right\vert e^{-ikx_{0}}}{%
\left\vert {(}\Delta_{+}{-}NU_{0}{+}i\kappa{)(}\Delta_{-}{-}NU_{0}{+}i\kappa{%
)-}N^{2}U_{0}^{2}\left\vert R_{-}\right\vert {^{2}}\right\vert ^{2}}.
\label{eq:intf}
\end{equation}

Here we see that the better the atomic localization $\left\vert
R_{-}\right\vert $ the more backscattering and stronger modulation of the
intracavity intensity we get. As a second important piece of information
this result allows us now to determine the spatial shift $\alpha$ of the
cavity field intensity maxima with respect to $x_{0}$ by help of Eq.\ref%
{intensity}. Since $U_{0}=-\left\vert U{_{0}}\right\vert =\left\vert U{_{0}}%
\right\vert e^{i\pi}$ we define the relative shift $\delta_{x}=\phi/(2k)$ of
the field intensity maxima with respect to $x_{0}$ by

\begin{equation}
\Delta_{-}-NU_{0}+i\kappa=\sqrt{(\Delta_{-}-NU_{0})^{2}+\kappa^{2}}
e^{i(\phi-\pi)}.
\end{equation}
This gives $\alpha_{\pm}=\phi-2kx_{0}=-2k(x_{0}-\phi/(2k))=-2k(x_{0}-%
\delta_{x})$, where the value of $\phi$ is determined by (\ref{eq:intf}):
\[
\label{shifteq} \cos\phi=\frac{-\Delta_{-}+NU_{0}}{\sqrt{%
(\Delta_{-}-NU_{0})^{2}+\kappa^{2}}},\;\;\sin\phi=\frac{-\kappa}{\sqrt{%
(\Delta_{-}-NU_{0})^{2}+\kappa^{2}}}.
\]

As the maxima of the light field intensity determine the minima of the
optical potential Eq. \ref{optpot} (stable equilibrium points), we see
immediately that field minima induced by the density distribution of the gas
do not coincide with the peak density positions $x_{0}$ but are shifted by $%
\delta_{x}$. Hence we obviously cannot find a self consistent time
independent configuration of particles and field as for the standingwave case%
\cite{Asboth05}. It is very interesting to note though, that the magnitude
of the shift only depends on the total particle number via $NU_{0}$ and
cavity parameters, but not on the pump strength $\eta$ or the precise form
of the atomic distribution beyond the average $\left\langle
e^{2ikx_{0}}\right\rangle $.

Obviously the above calculated shift of the potential implies an average
force on the cloud and an acceleration of the particles center of mass
position. This result can also be derived from basic momentum conservation
arguments as follows: since all photons in the backscattered mode have lead
to a deposition of $2\hbar k$ of momentum in the cloud, the gas is
accelerated, whenever a backscattered field is present. As a consequence of
this acceleration, $R_{-}$ will be time dependent in general. Hence in
contrast to atomic selforganization in a standing wave cavity field\cite%
{selforgEPL}, we cannot expect the system to reach a time independent self
consistent atom-field steady state. Nevertheless, as we will see below,
quasistationary self consistent atom field distributions can still form, if
this acceleration is not too fast.

\subsection{Quasistationary accelerated atomic distribution}

As argued above the coupled atom field dynamics produces an acceleration of
the particles center of mass. As long as this acceleration is not too fast
the corresponding periodic field intensity distribution will almost
instantaneously follow the particles center of mass motion. Hence the
particles see only a slowly time varying optical potential relative to their
center of mass position. This should allow the particle distribution to
adjust to this accelerating potential. In order to get some quantitative
insight here we will now transform our equations to a suitable \emph{%
accelerated} reference. Primarily this introduces an \emph{inertial force}
(acceleration): $F_{in}=mg$ into the atomic equations of motion, which can
be accounted by adding a term $V_{in}=-mgx$ to the effective atomic
potential. Obviously a suitable choice of $g$ requires to shift the periodic
local potential minima to coincide with the peaks of the atomic position
distribution. This allows subsequently to find an approximate stationary
atomic distribution with respect to the accelerated potential.

Of course the transformation to this frame is reflected in the field
equations, where the detunings are now time-dependent quantities induced by
the time dependent Doppler shifts of the cavity modes(\ref{modederiv}).
Nevertheless, as long as the acceleration is not too fast on the timescale
of the field evolution $\kappa$, the fields will still adiabatically follow
changes of atomic velocity and distribution. Lets consider this in a bit
more detail now. Omitting the constants parts, the potential in the
accelerated frame reads:

\begin{equation}
V(x)=-\left\vert U_{0}\right\vert \left\vert a_{+}a_{-}^{\ast}\right\vert
\cos(2k[x-(x_{0}-\frac{\phi}{2k})])-mgx,   \label{Vx}
\end{equation}
where we choose $g$ in a way, that the peaks of the atomic density
distribution $x_{0}+n\lambda/2$ coincide with local potential minima, i.e.
we want ($d/dx)V(x)=0$ at $x=x_{0}+n\lambda/2$. This requires
\begin{equation}
-2k\left\vert U_{0}\right\vert \left\vert a_{+}a{_{-}^{\ast}}\right\vert
\sin\phi+mg=0.
\end{equation}
Substituting the steady state expressions for $\sin\phi$ and for $%
a_{+}a_{-}^{\ast}$, we then get
\begin{equation}
g=-\left\vert {\eta^{2}}\right\vert \frac{2k}{m}\frac{N\left\vert
U_{0}\right\vert ^{2}\left\vert R_{-}\right\vert \kappa}{\left\vert {(}%
\Delta_{+}{-}NU_{0}{+}i\kappa{)(}\Delta_{-}{-}NU_{0}{+}i\kappa{)-}%
N^{2}U_{0}^{2}\left\vert R_{-}\right\vert {^{2}}\right\vert ^{2}}.
\label{accel}
\end{equation}

The resulting acceleration $g$ thus is \emph{negative} and depends on the
magnitude of the atomic localization parameter $R_{-}$ . This is consistent
with our expectations, since the mode $e^{2ikx}$ is pumped, and the
scattering converts photons travelling the negative $x$-direction to photons
in the positive direction and we have
\begin{equation}
mg=-4\kappa\left\vert a_{-}^{\ast}a_{-}\right\vert \frac{k}{N}.
\label{accel2}
\end{equation}

Besides providing the basis for the choice of an accelerated frame, Eqs.\ref%
{accel}, \ref{accel2} are quite interesting and useful results on their own.
While for small $N$ the acceleration linearly grows with $N$ showing the
superradiant $N^{2}$ enhancement of backscattering, for large enough $N$ it
decays like $1/N$. Similarly for large $\kappa$ (bad cavity) and a nearly
flat distribution $R\approx0$ acceleration vanishes. In these limits we can
expect the distribution to have enough time to equilibrate to a
quasistationary distribution in the accelerated frame.

Note that the acceleration for not too large $N$ depends on $|U_{0}^{2}|$
and thus on the inverse square of the particle-field detuning as it is the
case for near resonant radiation pressure and spontaneous emission. Hence at
first sight this does not look favorable for molecules. However, the
collective enhancement factor $N$ in the force will not show up in
spontaneous emission, which allows to strongly increase the force per
spontaneously scattered photon in the resonators.

Having now found a proper frame as a next step, we will calculate a
corresponding atomic distribution. For this we adapt the ideas developed for
the standing wave cavity case \cite{Asboth05} and simply use a thermodynamic
equilibrium canonical density distribution for the particles. Hence the
particle distribution is approximated by
\begin{equation}
\rho(x)=\frac{1}{Z}\exp\left[ -V(x)/k_{B}T\right] ,   \label{rho1}
\end{equation}
where the normalization is given by the partition function
\begin{equation}
Z=\int\exp\left[ -V(x)/k_{B}T\right] dx.
\end{equation}

Note that the effective potential now contains the acceleration terms $mgx$
and this procedure only makes sense as long as $V(x)$ has periodic local
minima of sufficient depth. This is definitely be the case for small $g$ as
it occurs for an initial nearly homogeneous particle distribution, where $%
R_{-}$ is close to zero. Hence we can at least use this formalism to study
the initial phase and threshold conditions of the selforganization process.

\subsection{Selforganization threshold}

As we have seen above the cavity field and inertial force from the
acceleration $g$ determine an effective combined potential $V(x)$ for the
accelerated particles for which we get a corresponding stationary
distribution from Eq.\ref{rho1}. This modified density distribution
corresponds to a new bunching parameter $R$ and a new optical potential. By
subsequent application of Eqs. \ref{eq:intf}, \ref{Vx} this procedure can be
iterated and in many cases a self consistent solution is obtained for $%
\rho(x)$ and $V(x)$ after a few iterations. Note that in order to get a
meaningful result, we have to restrict ourselves to a single potential well
(local potential minimum) to determine the density distribution $\rho(x)$
and assume that all other wells to have a similar number of particles. This
can be physically motivated from the fact that redistribution between
different wells will occur on a much longer time scale as internal
thermalization. Note the we do not account for direct interparticle
interactions here, which would occur for very high local densities.

There are now several decisive aspects that can be learned about the physics
of our problem from this mathematical iteration procedure. We will start
here with the most simple but central question, whether a homogeneous
initial particle distribution at rest within the resonator is unstable
against small perturbations. This amounts to a CARL type setup, where
particles, which are originally evenly distributed are collectively
accelerated by superradiant scattering\cite{Tuebingen}. In the following we
will first determine the minimal pumping threshold required to initiate such
superradiant backscattering via self ordering. In a second step we will then
check, whether a quasistationary accelerated ordered phase can develop out
of these initial fluctuations.

Starting from an initial flat particle distribution, we obviously have $%
R_{-}=0$ and the atoms generate no coherent backscattered field. Hence the
corresponding average acceleration is zero at the beginning and we can start
in laboratory rest frame for the dynamics. Superradiant backscattering and
collective forces in this case can only build up, if this solution is
unstable against small density fluctuations with period $\lambda/2$. Hence
we add an infinitesimal perturbation to our uniform distribution $\rho(x)=%
\frac {1}{\lambda}$ in the form:

\begin{equation}
\rho^{(0)}(x)=\frac{1}{\lambda}(1+\epsilon r(x)).   \label{rho0}
\end{equation}
$r(x)$ in general could be any periodic function
\begin{equation}
r(x)=\sum\limits_{m=1}^{\infty}(A_{m}\cos(mkx)+B_{m}\sin(mkx)),\text{ }%
\sum\limits_{m=1}^{\infty}A_{m}^{2}+B_{m}^{2}=1
\end{equation}
and see whether this perturbation is amplified during an iteration cycle.
Note that within the ring cavity geometry the only perturbation that
contributes to $R_{-}$ is that for $m=2$, i.e. we can limit ourselves to $%
r(x)=A_{2}\cos(2k(x-x_{0}))$+$B_{2}\sin(2k(x-x_{0}))$ as a perturbation. Due
to translational invariance of the problem we can also set $B_{2}=0$. Hence
the perturbation simply leads to $R_{-}=\epsilon A_{2}/2$ for the mode
coupling parameter.

As a first step in the iteration we will now use this $R_{-}$ to determine
the corresponding fields and the optical potential. As shown above in Eq.\ref%
{Vx} the optical potential is shifted with respect to $x_{0}$ by an amount
independent of $A_{2}$. In the second step we now choose the acceleration $g$
in a way described above to pin the effective potential minimum to $x_{0}$
and apply Eqs. \ref{Vx} and \ref{rho1} to find the corresponding thermal
density distribution for this potential. Of course this distribution now
shows a modulation induced by the optical potential which we have to compare
with the magnitude of our initial perturbation.

As $\epsilon$ is small we do this analytically by expanding $\rho(x)$ in Eq.%
\ref{rho1} to lowest order in $\epsilon$ to find:

\begin{align}
\rho^{(1)}(x) & =(1-\frac{V(x)}{k_{B}T})/\lambda= \\
& =\frac{1}{\lambda}(1-\frac{\epsilon A_{2}}{k_{B}T}\frac{%
NU_{0}^{2}\left\vert {\eta}\right\vert {^{2}}}{\left\vert {(}%
\Delta_{-}-NU_{0}-i\kappa{)}((\Delta_{+}-NU_{0})^{2}+\kappa^{2})\right\vert }%
\left[ \cos(2k(x-x_{0})\right] +O(\epsilon^{2})
\end{align}

The central question now is, whether the density perturbation has been
enlarged or diminished in this iteration cycle. For a flat distribution to
be stable, the first order correction term in $\epsilon$ must be less than $%
A_{2}$. Thus stability requires:

\begin{equation}
\eta_{thresh}>\sqrt{\left\vert \frac{k_{B}T \sqrt{(\Delta_{-}-NU_{0})^{2}+%
\kappa^{2}}\;((\Delta_{+}-NU_{0})^{2}+\kappa^{2})}{NU_{0}^{2}}\right\vert }.
\label{threshv}
\end{equation}

This formula is one of the central results of this work, which will discuss
in more detail at different examples in the following. It allows to clearly
determine the experimental conditions for the cavity, initial particle phase
space density and laser power required for such setup. We will later also
substantiate the key properties of this formula by direct multi particle
simulations.

\section{Results and discussion}

\subsection{Threshold condition and acceleration of a gas at rest}

To investigate the underlying physics in more detail let us first consider a
cloud at rest $v=0$, i.e $\Delta_{+}=\Delta_{-}=\Delta_{c}$. In this case
the threshold condition reads:

\begin{equation}
\eta_{thresh}>\sqrt{\left\vert \frac{k_{B}T\left(
(\Delta_{c}-NU_{0})^{2}+\kappa^{2}\right) ^{3/2}}{NU_{0}^{2}}\right\vert }.
\label{etagen}
\end{equation}

Obviously a flat distribution always gets unstable for sufficient pump
power. One needs less pump intensity the more particles $N$ one has and the
stronger they are coupled to cavity mode (bigger modulus of $U_{0}$). The
required pump intensity scales linearly with the initial temperature, which
gives some important restriction for useful molecular sources in particular.
Of course the validity of the threshold formula has its limitations as the
weak saturation and two level approximation for the particles eventually
will break down at certain pump intensities.

Note that this stability threshold for a flat distribution attains a minimum
value, when $\Delta_{c}=NU_{0}$, where it simplifies to

\begin{equation}
\eta_{thresh}>\sqrt{\frac{k_{B}T\text{ }\kappa^{3}}{N U_{0}^{2}}}.
\label{etamin}
\end{equation}

Of course $\Delta_{c}=NU_{0}$ is just the resonance condition for the cavity
mode shifted by the refractive index of the particles and hence gives a
maximum of the intracavity photon number. As we will see later from Eq.\ref%
{shifteq} this is a rather tricky parameter choice. In this case the
corresponding shift of the optical potential minima with respect to the
atomic center of mass just corresponds to $\pi/4$. Here the particles sit on
the position of maximum slope and the optical dipole force has maximum as
well. However, as the curvature of the potential is zero at this point the
effective potential in the accelerated frame has no local minimum and one
cannot expect the formation of a self consistent atomic distribution of the
form of Eq.\ref{rho1}.

\begin{figure}[ptb]
\includegraphics[height=60mm]{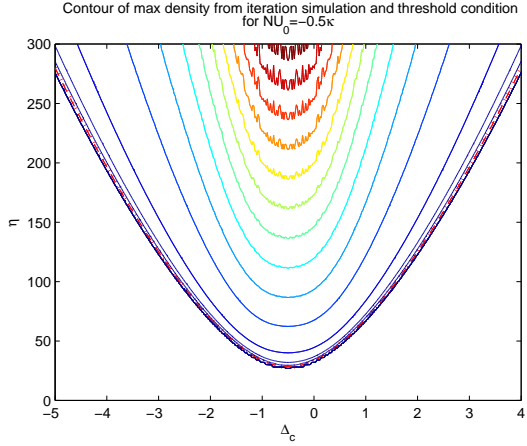} 
\caption{Contour lines of the maximum of the selfconsistent atom density
distribution for a cloud at rest as a function of detuning $\protect\delta%
_{c}$ and pump amplitude $\protect\eta$. The analytic formula for the the
selforganization threshold is indicated by the red dashed line. Parameters
are $N=200,U_{0}=-1/400,\protect\kappa=1.$}
\label{NUhalf}
\end{figure}

\begin{figure}[ptb]
\includegraphics[height=60mm]{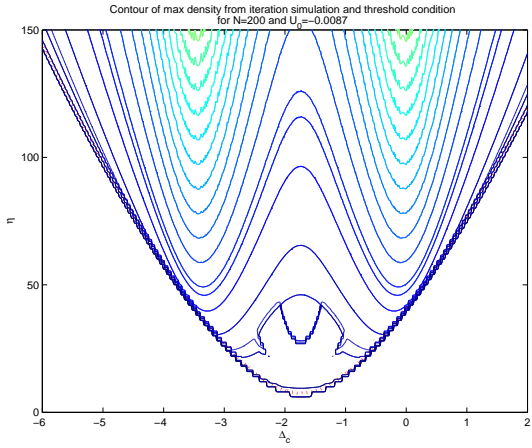} 
\caption{Same as Fig. \protect\ref{NUhalf} for $N=200,U_{0}=-1/115,\protect%
\kappa=1.$}
\label{N200}
\end{figure}

\begin{figure}[ptb]
\includegraphics[height=60mm]{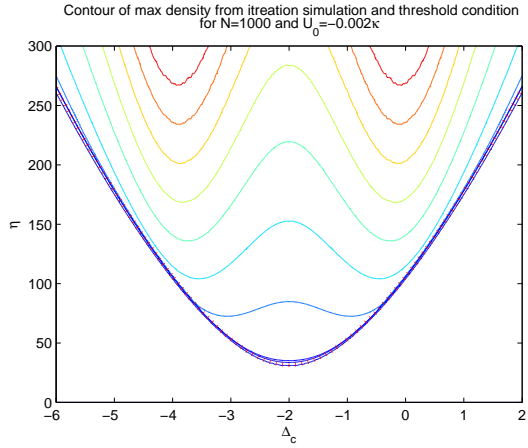} 
\caption{Same as Fig. \protect\ref{NUhalf} for $N=1000,U_{0}=-1/375,\protect%
\kappa=1.$}
\label{N1000}
\end{figure}

In the following we will investigate the validity of this threshold
condition numerically by iterating equations \ref{Vx},\ref{rho1} for a
selfconsistent atom-field configuration. We see that the iteration results
agree nicely with the analytic condition of Eq.\ref{etamin}. As shown in
Figs. \ref{NUhalf} \ref{N200}, \ref{N1000} and \ref{N200} we find a
numerical instability in agreement with Eq.\ref{etamin} with the minimum
threshold at $\Delta _{c}=NU_{0}$. \ The onset of self ordering contour is
readily seen to closely follow the red dashed line representing the analytic
threshold condition Eq.\ref{etagen}. \ $N=200,U_{0}=-1/115,\kappa=1.$

Let us now further substantiate these predictions for some special parameter
choices by a real space dynamical solution of the equation of motion of N
particles and the field modes. For this we use a semiclassical dynamical
model as developed in ref. \cite{Domokos03}, where the atoms are modeled by
polarizable point particles and the fields are amplitudes are represented by
coherent states. This model has already been successfully applied to
simulate cavity enhanced laser cooling and atomic selforganization in a
standing wave cavity geometry\cite{Asboth05}. Starting from a flat thermal
distribution atomic distribution and with zero field amplitude, we then
simply monitor the coupled atom-field evolution for a set of typical
parameters. Due to the large number of variables we have to restrict
ourselves to monitor only some characteristic observable like the evolution
of the total kinetic energy, the magnitude of the bunching or the time
evolution of the center of mass of the cloud. This is shown in Fig.\ref%
{energy} where we compare contour lines the kinetic energy gain of the gas
after a time of $\kappa t=60(check)$ with the analytic threshold formula
from above. Indeed selfordering and significant acceleration of the
particles is found only above the analytic threshold line.

\begin{figure}[ptb]
\includegraphics[height=60mm]{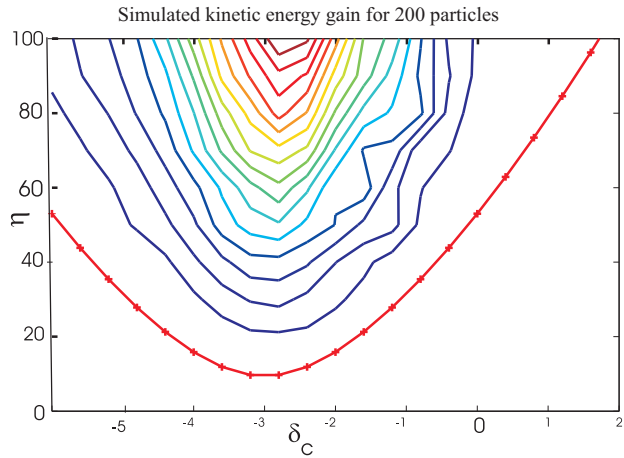}
\caption{Contour lines of kinetic energy gain of N=200 particles with
average initial velocity zero and average kinetic energy $\hbar\protect\kappa
$ and $U_{0}=-0.015\protect\kappa$ within a fixed time interval $\protect%
\kappa t=60$. }
\label{energy}
\end{figure}

\subsection{Quasistationary accelerated particle distribution}

As we now found a condition for the breakdown of stability of a homogeneous
distribution, the next big question now is what happens in this instable
regime? What kind of dynamics can we expect in general and under what
conditions can we expect selforganization and an ordered accelerated
particle distribution?

Obviously the backscattered field amplitude gets nonzero and the center of
mass of the cloud gets accelerated. As a first guess for the further
evolution we will now simply iterate the above loop further and see under
which conditions a self consistent accelerated atom field distribution could
possibly form. As we have no reliable time scales for the atomic
thermalization process available, we will simply assume that it is faster
than the time scale of significant acceleration of the cloud so that we keep
the average velocity constant in these iterations.

Experimentally and in numerical simulations it has been found that in
particular the regime $NU_{0}\geq\kappa$ has the biggest potential for a
self ordered phase and interesting dynamics, so we will primarily focus on
this regime\cite{Asboth05}.

\begin{figure}[ptbp]
\includegraphics[height=60mm]{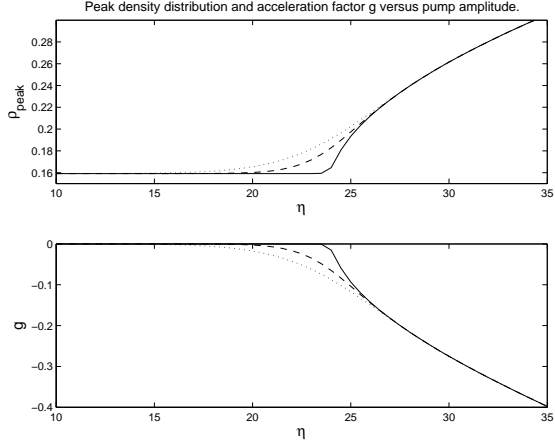} 
\caption{Maximum atomic density as a function of pump amplitude for three
different iteration numbers of 5 for dotted line, 10 for dashed line, and
100 for solid line.}
\label{numiter}
\end{figure}

In general for a wide range of parameters a self consistent accelerated atom
field distribution can be found at a sufficient pump strength. A plot of the
density distribution below, at, and above pump threshold is shown in Fig \ref%
{position_distribution}.
\begin{figure}[ptb]
\includegraphics[height=60mm]{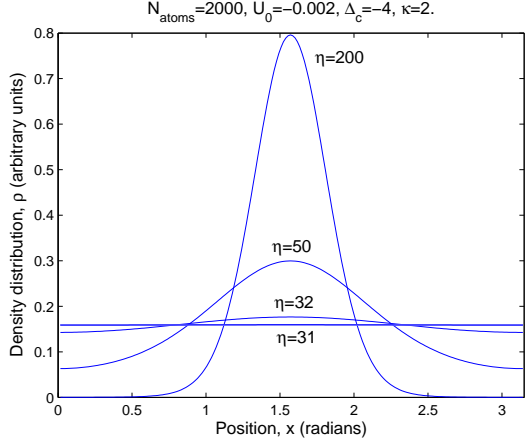} 
\caption{Self consistent atomic density distribution below and above
threshold for $N=2000$ atoms for $U_{0}=-0.001\protect\kappa$ and $\protect%
\delta_{c}=-2\protect\kappa$. The analytic threshold result for this set of
parameters is $\protect\eta_{thresh}=31.6.$}
\label{position_distribution}
\end{figure}
Actually the iterations show, that the onset of self ordering depends non
trivially on all the parameters and clear indications can already be seen
even after only about five iterations. Typical numerical results can be
visualized by plotting the peak atomic density versus pump strength for 5,
10 and 100 iterations as in Fig \ref{numiter}. There are no significant
differences in convergence between 50 and 100 iterations. This peak density
is a good indicator of atomic ordering, which due to translational
invariance may form at any location between the mirrors. Alternatively the
self consistent backscattered intensity turns out to be a good indicator for
the onset of self ordering as a function of pump strength. From this
quantity we can also determine the average force on the cloud.

The self ordering threshold can be located in Fig \ref{numiter} or from the
contour plots such as Fig \ref{N200} or \ref{N1000} via the transition from
the flat homogeneously distributed background to the exponential increase of
local particle density. The transition is easy to see by eye. Numerically we
have to select some value of particle density larger than the homogenous
density and above the "noise background" as an indicator of
selforganization. In this way we can generate contour lines of the maximum
density of the self consistent particle density. This allows a good
comparison with the analytic form for the pumping threshold as a function of
other parameters such as detuning. A plot of the density distribution below,
at, and above threshold is shown in Fig \ref{N200}. The key feature to note
in Figs \ref{N200} and \ref{N1000} is that in accordance with the analytical
results derived above, the minimum threshold which occurs, when $%
\Delta_{c}=NU_{0}$, with a symmetric increase around this minimum. In seems
that for a large range of parameters the instability of the flat
distribution is connected with the existence of an ordered accelerated phase
which is particularly stable for $\Delta_{c}<NU_{0}$.

\begin{figure}[ptb]
\includegraphics[height=60mm]{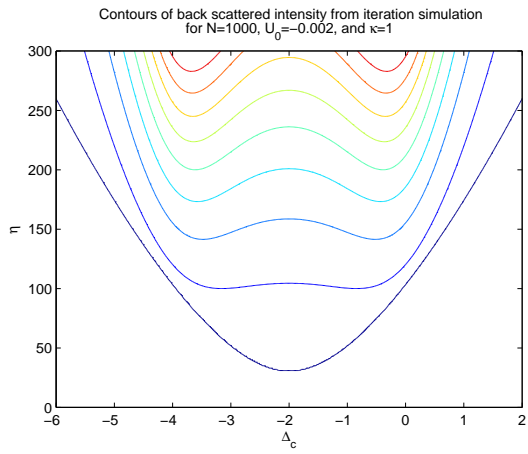} 
\caption{Intensity of backscattered field mode as a function of detuning $%
\protect\delta_{c}$ and pump amplitude $\protect\eta$ for $N=1000$ atoms $%
U_{0}=-0.002 \protect\kappa$ }
\label{back1000}
\end{figure}

A decrease of cavity damping leads to a sharper resonance of backscattered
light at certain values of cavity detuning. We can estimate the value of
cavity detuning which leads to the maximum backscattering (and average
force) while holding other parameters fixed by taking the derivative of $%
\left\vert a_{-}\right\vert ^{2}$,
\begin{equation}
\Delta_{c\text{ }Fmax}=NU_{0}\pm\sqrt{N^{2}U_{0}^{2}\left\vert
R_{-}\right\vert ^{2}-\kappa^{2}}   \label{backscatter max}
\end{equation}
\ Recall the bunching parameter approaches unity here. \ This result is only
useful for small damping, $\kappa<NU_{0},$ where there is a distinct, sharp
maxima of $\left\vert a_{-}\right\vert ^{2}$ in the numerical results at $%
\Delta_{c}=0$ and $\Delta_{c}=2NU_{0}$\ (with very little backscattering at $%
\Delta_{c}=NU_{0}).$\ \ Thus as seen in Fig. \ref{back1000} there are
effectively these two resonant frequency of the cavity, with a larger Q
leading to more backscattered light, but over a narrower $\Delta_{c}$
frequency interval. \

Alternatively, for larger values of damping, $\kappa>NU_{0},$ the numerical
results for backscattering show a maximum value at $\Delta _{c}=NU_{0}$
which decreases slowly away from this value of detuning. The limit on the
overall range of cavity detuning which still permits self ordering of the
particles is set by the pump. \ The threshold pump value given in Eqs \ref%
{etamin} and \ref{etagen} should be exceeded to observe these backscattering
dependencies on detuning.

To get some quantitative comparison with the selfconsistent atomic
distributions calculated by the above iteration procedure and direct
numerical simulations, we now also depict a contour plot the maximum of the
atomic density distribution of a cloud of $N=1000$ atoms $NU_{0}=-2\kappa$
after a given time interval $\Delta t=xx/\kappa$ . We see that clear signs
of selforganization appear when we cross the theoretically found threshold
condition depicted as the line with crosses as a function of detuning and
pump strength. We also see that crossing the threshold and ordering is
accompanied by acceleration as expected.

\begin{figure}[ptb]
\includegraphics[height=60mm]{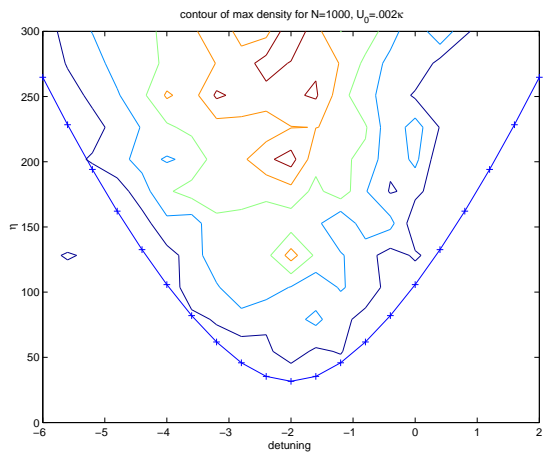} 
\caption{maximum of atomic density distribution }
\label{energypar}
\end{figure}

\subsection{Threshold conditions and deceleration prospects for a moving
thermal cloud}

Having investigated the dynamics of a gas at rest in some detail, we now
come back to one of our original central goals and check the possibility of
trapping and cooling molecules from a beam by superradiant scattering. In a
typical setup one would send a molecular beam through the cavity in a small
angle almost parallel to the ring cavity axis and inject a strong counter
propagating coherent field. In such a setup the molecules are initially
moving quite fast (up to \symbol{126}100 m/s) with constant spatial density.
The central question now is for which phase space densities, velocities and
pump amplitudes superradiant scattering will appear and can be used slow
down and trap the molecules.

In general using the rest frame of the beam the equations are the same as
above except for different detunings $\Delta_{\pm}=\Delta_{c}\pm kv.$ for
both cavity modes. Hence one can expect a similar threshold behavior.
Naturally for a large value of the ratio of $kv/\kappa$ the pump and the
backscattered field cannot be on resonance simultaneously. This will
increase the threshold or even make self ordering not be possible. In
principle on has the two possibilities of resonant pumping but a non
resonant backscattered field or off resonant pumping to get resonant
enhanced of the backscattered field. \ The threshold for self ordering is
first calculated assuming the cloud velocity is constant throughout the
iteration. \ We can compare this result to the threshold obtained using the
accelerated frame where the velocity is updated throughout the iteration. \

\begin{figure}[ptb]
\includegraphics[height=60mm]{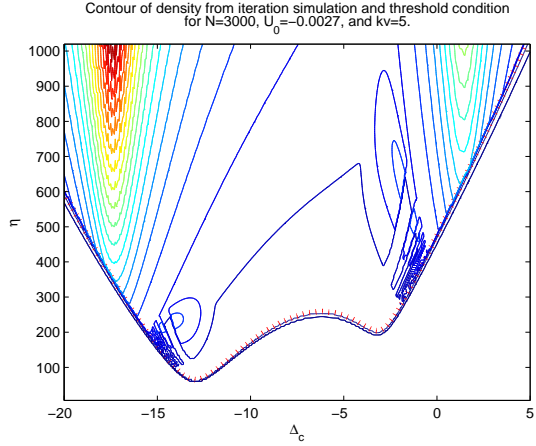} 
\caption{Contour lines of maximum of the self consistent atomic density for $%
N=3000$ atoms for $U_{0}=-0.0017\protect\kappa$ and $k v=5\protect\kappa$.
For comparison we also show the analytical expression for the
selforganization threshold amplitude multiplied by 3 (dotted line).}
\label{N3000}
\end{figure}

The most notable feature of the numerical results as seen in Fig. \ref{N3000}
is the asymmetry in the pumping threshold \ with detuning as compared to the
case with particles at rest. \

Additionally, the analytic stability analysis leads to a first order
stability equation for optical pumping threshold given by Eq. \ref{threshv}

For $NU_{0}>\kappa$ there is once again excellent agreement between the
numerical and analytic results for pumping threshold as shown in Fig. \ref%
{N3000} with only about a two percent difference in threshold values. \
Trapping molecular beams with relatively large velocities is possible with a
large pump and large cavity decay rate. \ In this region where $%
NU_{0}<\kappa $, the analytic and numerical results for threshold are less
in agreement, although the results have the same functional form as $kv$
increases.The backscattered intensity as a function of pumping versus
velocity of the moving cloud is shown in Fig. \ref{back1000}.


\subsection{Numerical multiparticle simulations of cold beam stopping}

As a final check and demonstration of the efficiency of the setup we will
now present some results on the possibility to stop a fast beam in such a
setup again using real space simulations. In this case we start with beam of
N=3000 particles in a thermal distribution of temperature $kT=\hbar\kappa$
but with additional average momentum of $p_{z}=-2000\hbar k$ per particle in
the negative z-direction counter propagating the pump. As a measure of the
efficiency of the scheme we first plot the amount of average kinetic energy
extracted from the atoms by the fields in a time interval $\kappa t=60$ as a
function of pump strength $\eta$ and pump detuning $\Delta_{c}$.

\begin{figure}[ptb]
\label{energy_loss} \includegraphics[width=6cm] {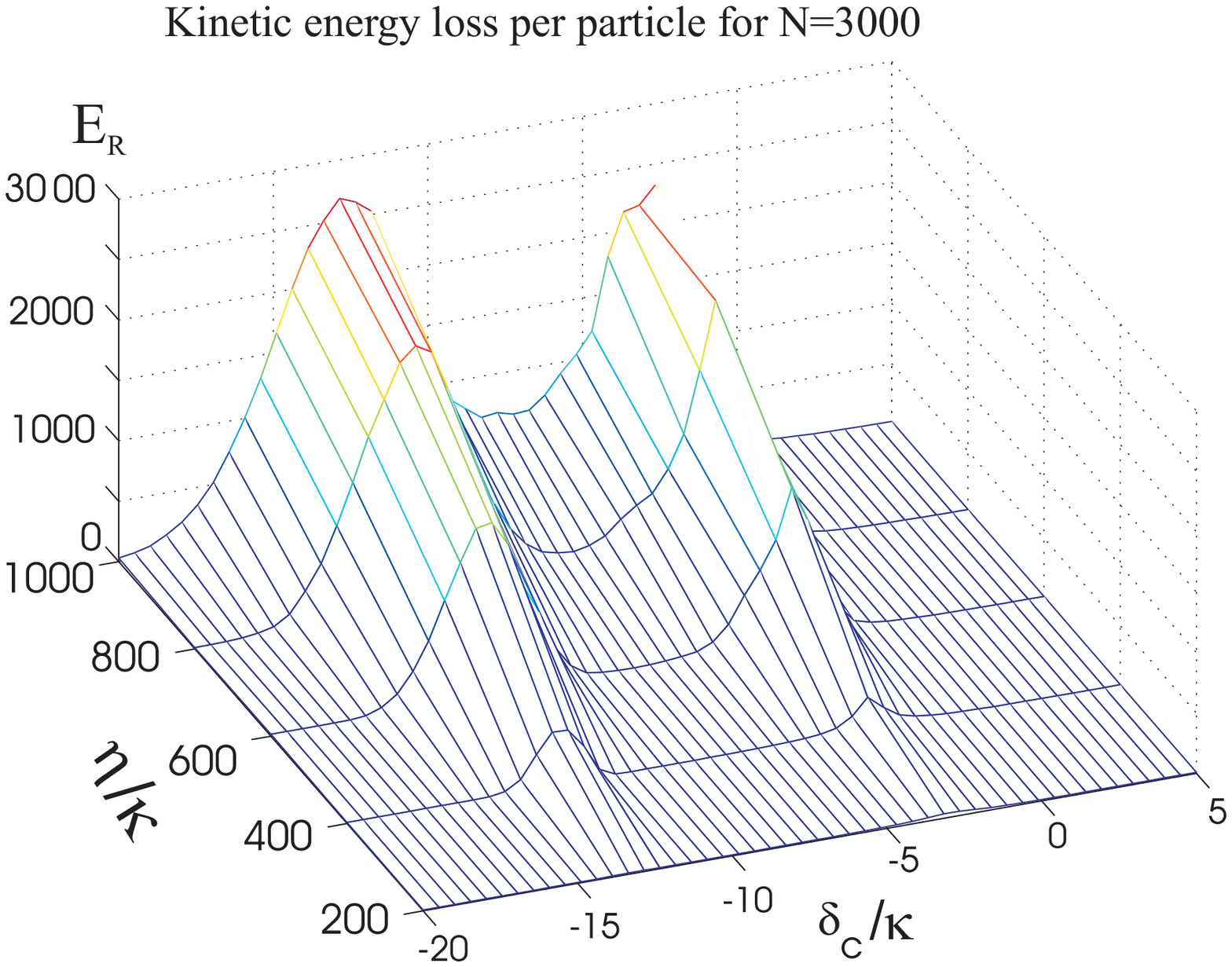} %
\includegraphics[width=6cm] {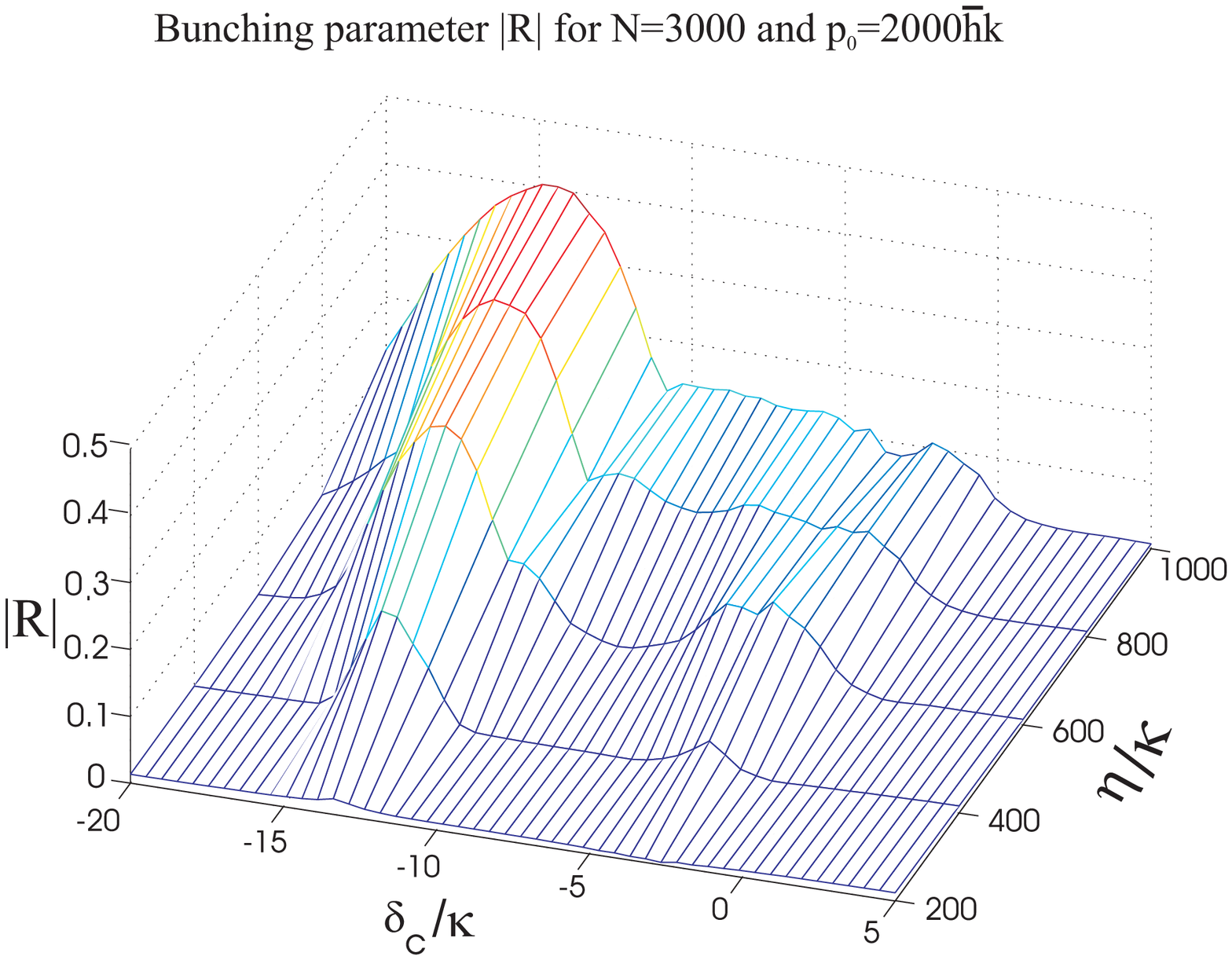} %
\caption{(a) Kinetic energy loss per particle within a time interval of $%
\protect\kappa t=60$ as a function of detuning $\Delta_{c}$ and pump
amplitude $\protect\eta$ for $N=3000$ atoms counterpropagating the cavity
axis with initial average momentum of $p_{0}=2000\hbar\protect\kappa$ and
temperature $kT_{0}=\hbar\protect\kappa$ and (b) average bunching parameter
during this time for the same parameters. }
\end{figure}

Obviously we get efficient deceleration of the cloud when either the pump
mode directly or the reflected light field is in resonance with the
effective cavity eigenfrequencies\cite{Hemmerich}. When we plot the time
averaged bunching parameter we see that efficient energy transfer is
associated with atomic selfordering as analytically predicted. As the atomic
bunching is more efficient for negative detunings we get a more efficient
energy extraction in this limit.

\begin{figure}[ptb]
\label{bunchcontour} \includegraphics[width=6cm]
{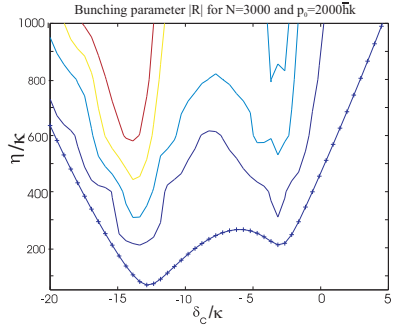}
\caption{Contour lines of time averaged bunching parameter $|R|$ found by
numerical simulation as function of detuning $\protect\delta_{c}$ and pump
amplitude $eta$ for parameters above. Again for reference we show the
analytical result for the selforganization threshold (x3) }
\end{figure}

These results compare qualitatively quite well with the analytic prediction
of Eq.\ref{threshv} as shown in Fig.\ref{bunchcontour} where we compare a
contour plot of bunching parameter with the analytic threshold formula for
the selforganization threshold multiplied by 3 (line with crosses).

\section{Conclusions}

We studied the conditions to stop and trap a fast molecular beam by
help of a single side pumped high Q ring cavity geometry in a CARL
setup. The analytical results for kinetic energy extraction from a
self consistent stationary solution of the coupled particle field
equations were checked by numerical simulations. In the limit of
large detuning inversion is negligible and coherent light scattering
dominates. As the atom field interaction strength per particle is
small in this regime we need collective enhancement of the Raleigh
scattering blue shifted with respect to the excitation light to do
the job. This can be achieved in a regime where the particles
arrange in a periodic density lattice and form a moving Bragg
mirror.

Finding the effective optical potential in an accelerated frame and applying
a linear stability analysis allows to derive the analytic threshold for the laser
pump power, above which an initial homogenous particle distribution gets
unstable against small density fluctuations. We see that the available phase
space density at the beginning determines the required threshold laser
intensity. Above this threshold a suitable choice of detunings causes the
interference of the injected and collectively backscattered light fields to
establish a periodic particle density acting like a moving Bragg mirror,
which yields a time dependent coupling of the counter propagating fields
within the resonator. By using a mean-field analysis the self consistent
atom-field distribution is found by an iteratively which allows
to quantitatively predict the maximal acceleration and localization of the cloud.

In a second step we checked these predictions by simulating the microscopic
equations of motions of the particles for finite ensembles. These
simulations show surprisingly good qualitative and quantitative agreement
with the analytic threshold result and acceleration predictions within the
regime, where the analytical model predicts selforganization. Outside this
regime, where the effective potential in the accelerated frame shows no
periodic local minima the simulations differ from the analytic predictions.
This can be traced to cavity induced heating processes in this regime as
well as remnant radiation pressure not accounted for in the analytic model.
In any case the analytically found threshold formula for the selforganization process
provides a solid criterion for the prospects of a corresponding experimental setup.

As a bottom line trapping and stopping of a fast molecular beam in a ring
cavity seems experimentally feasible with current technology, provided one
can find a strong enough pump laser at a frequency where the particles are
polarizable but sufficiently transparent. The higher the initial phase space
density of the particles, the faster this process will happen and the
stronger are the deceleration forces and the suppression of spontaneous
emission. In favorable cases this could be combined with cavity cooling
in the particles rest frame as well\cite{Domokos03}.

This work could be extended by continuing the work of {Asb\'{o}th} who has
used the mean field analysis to find self-organization thresholds in
standing wave cavities, but now analyzed with transverse mode degeneracy. \
In particular, Vuleti{\'{c}'s} group has found the degeneracy of a confocal
resonator significantly enhances the cooling efficiency. \ It would be
useful to examine such threshold behavior in this highly degenerate regime.

\bigskip

\section*{Acknowledgments}

This work was supported by the Air Force Office of Scientific
Research and the Institute for Information Technology Applications and the Austrian Science Fund
(FWF) by grants S1512 and P17709.

\end{document}